\documentclass[aps,pra,twocolumn,amsmath,amssymb,superscriptaddress,longbibliography,floatfix]{revtex4-1}

\newcommand{\bea}{\begin{eqnarray}}
\newcommand{\eea}{\end{eqnarray}}
\newcommand{\beq}{\begin{equation}}
\newcommand{\eeq}{\end{equation}}

\usepackage[urlcolor=blue,colorlinks=true,citecolor=blue,linkcolor=blue,pdfstartview={FitH},bookmarks=false]{hyperref}
\usepackage{graphicx}
\usepackage{longtable}
\usepackage{epsfig}
\usepackage{dcolumn}
\usepackage{bm}
\usepackage{amssymb}
\usepackage{multirow}
\usepackage{times,color}
\usepackage{hyperref}
\newcommand{\cblue}{\color{black}}

\begin{document}

\title{Interplay between Zhang-Rice singlets and high-spin states in
a model for doped NiO$_2$ planes}

\author{Tharathep Plienbumrung}
\affiliation{\mbox{Institute for Functional Matter and Quantum Technologies,
University of Stuttgart, Pfaffenwaldring 57, D-70550 Stuttgart, Germany}}
\affiliation{\mbox{Center for Integrated Quantum Science and Technology,
University of Stuttgart, Pfaffenwaldring 57, D-70550 Stuttgart, Germany}}

\author{Maria Daghofer}
\affiliation{\mbox{Institute for Functional Matter and Quantum Technologies,
University of Stuttgart, Pfaffenwaldring 57, D-70550 Stuttgart, Germany}}
\affiliation{\mbox{Center for Integrated Quantum Science and Technology,
University of Stuttgart, Pfaffenwaldring 57, D-70550 Stuttgart, Germany}}

\author{Andrzej M. Ole\'s$\,$}
\email{a.m.oles@fkf.mpi.de}
\affiliation{\mbox{Max Planck Institute for Solid State Research,
             Heisenbergstrasse 1, D-70569 Stuttgart, Germany} }
\affiliation{\mbox{Institute of Theoretical Physics, Jagiellonian University,
             Profesora Stanis\l{}awa \L{}ojasiewicza 11, PL-30348 Krak\'ow, Poland}}

\begin{abstract}
Superconductivity found in doped NdNiO$_2$ is puzzling as two local
symmetries of doped NiO$_2$ layers compete, with presumably far-reaching
implications for the involved mechanism: a cuprate-like regime with
Zhang-Rice singlets {\cblue is replaced by local triplet states at
realistic values of charge-transfer energy, which would suggest a rather
different superconductivity scenario from high-$T_c$ cuprates}.  We
address this competition by investigating Ni$_4$O$_8$ clusters with
periodic boundary conditions in the parameter range relevant for the
superconducting nickelates. With increasing value of charge-transfer
energy we observe upon hole doping the expected crossover from the
cuprate regime dominated by Zhang-Rice singlets to the local triplet
states. We find that smaller charge-transfer energy $\Delta$ is able to
drive this change of the ground state character when realistic values
for nickel-oxygen repulsion $U_{dp}$ are taken into account. For large
values of the charge-transfer energy, oxygen orbitals are less important
than in superconducting cuprates as their spectral weight is found only
at rather high excitation energies. However, a second Ni($3d$) orbital
can easily become relevant, with either the $xy$ or the $3z^2-r^2$
orbitals contributing in addition to the $x^2-y^2$ orbital {\cblue to
the formation of triplet states. In addition,} our result that
$U_{dp}$ (acting between Ni and O) favors onsite triplets implies that
correlation effects beyond purely onsite interactions should be taken
into account when obtaining effective two-band models.
\end{abstract}

\date{\today}

\maketitle


\section{Introduction}
\label{sec:int}

Two-dimensional (2D) nickelates such as LaNiO$_2$ have been
theoretically proposed long ago \cite{Ani99} as candidate materials for
unconventional superconductors, but only recently has superconductivity
been found in Sr-doped NdNiO$_2$ thin films \cite{Li19}. This discovery
could be classified as fulfilling the paradigm of high-$T_c$
superconductivity in systems similar to cuprates where both $e_g$
symmetries contribute at the Fermi surface \cite{Ole19}. Indeed,
nickelate heterostructures were considered to be the most promising
\cite{Han09,Dis15}, but it took another decade until the
superconductivity was found in quasi-2D films \cite{Li19}. Shortly
after the discovery of superconductivity in doped NdNiO$_2$, the search
for its mechanism began and it became clear that once again we have to
do with unconventional superconducting (SC) materials. But the
situation in the theory is somewhat similar to SC cuprates, where after
their discovery in 1986 \cite{Mul86}, the mechanism of high-$T_c$
superconductivity remains still puzzling \cite{Lee06,Kei15}.

Recently several groups calculated the electronic structure and tried
to understand the essential differences to cuprate physics. Two
remarkable differences between the CuO$_2$ planes of correlated
insulator La$_2$CuO$_4$ and the NiO$_2$ planes in NdNiO$_2$ became
clearly evident. First, the nickelates are 'self-doped' due to the
their rare-earth bands, so that an 'undoped' compound does not
correspond to an undoped NiO$_2$ layer~\cite{Zha20,Ban20}. {\cblue
Regardless of the level of self-doping realized in superconducting
NdNiO$_2$, we focus here on understanding the undoped NiO$_2$ layers,
which we regard as a kind of idealized 'parent compound'.}
The argument
for studying a NiO$_2$ plane is a certain effective decomposition into
a three-dimensional band including rare-earth states and a more
2D band of $x^2-y^2$ character~\cite{Hep20}. Second, the NiO$_2$
layers miss the apical oxygen ions present in the CuO$_2$ layers.
However, the NiO$_2$ plane of the infinite 2D layer in NdNiO$_2$ is
similar to the CuO$_2$ plane of CaCuO$_2$, where the apical oxygens are
missing as well. One might argue that the properties of these planes
would be then similar but they in fact reflect two different parameter
regimes of the charge-transfer model which describes them both.

\begin{figure}[t!]
\includegraphics[width=.99\columnwidth]{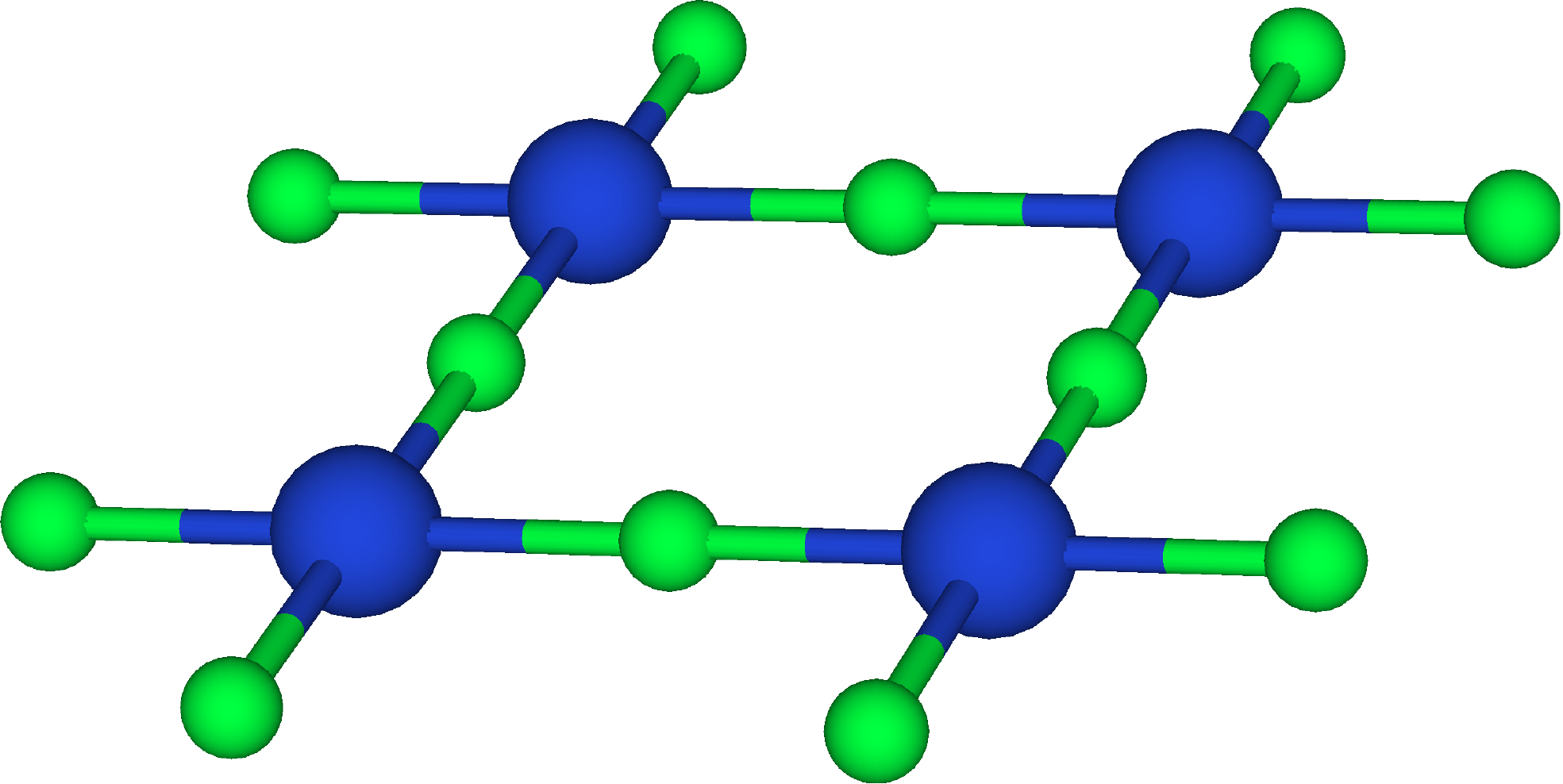}
\caption{Schematic picture of the 2D periodic cluster (NiO$_2)_4$ used
here to model the NiO$_2$ infinite plane. Blue and green balls stand
for Ni and O ions in the plane.}
\label{cluster}
\end{figure}

In spite of great similarity between CuO$_2$ and NiO$_2$ planes, there
are thus substantial differences in the electronic structure. While
La$_2$CuO$_4$ is a charge-transfer system, the NiO$_2$ planes in
NdNiO$_2$ have large charge-transfer energy $\Delta$ which indicates
more a Mott-Hubbard system \cite{Jia20}. This is also confirmed in other
works \cite{Lec20,LecPRX,Li20,Si20,Zha20,Adh20,Gei20} and thus Ni-O
hybridization plays here a less important role than in cuprates.
Arguments were also given that electronic structure in nickelate
superconductors could be reduced to a one-band Hubbard model
\cite{Kit20}. The undoped system has one hole in $x^2-y^2$ Ni orbital
which is the $|b_1\rangle$ symmetry state but Zhang-Rice (ZR) singlets
($S=0$) $|b_1L_{b_1}\rangle$ are much weaker in doped systems. These
states compete with high-spin ($S=1$) $|a_1b_1\rangle$ states when
doping increases \cite{Jia20}.

But perhaps the most important difference between the above two classes
of materials is that there is only one band which crosses the Fermi
level in cuprates, while two bands cross the Fermi level in nickelates
\cite{Ole19,Bot20,Adh20,Pet20}. This suggests that indeed both $e_g$
orbitals could contribute at finite Sr doping to the properties of
NdNiO$_2$. In the impurity model one finds therefore a transition from
the singlet to triplet local configuration at doped Ni$^{2+}$ ion
\cite{Jia20}, and we shall investigate here how this changes by going
to a more extended Ni-O system than a single NiO$_4$ unit, i.e., beyond
the impurity model. Furthermore, we show below that the transition to
the regime where high-spin states contribute depends on the value of
intersite Coulomb repulsion $U_{dp}$.

Another question which we want to address here is the nature of
high-spin states in doped materials. In a cuprate model the orbital
which is close to the top $x^2-y^2$ orbital is the second $e_g$ state
of $3z^2-r^2$ symmetry (called also $z^2$ below), and $S=1$ states could
form by hole doping \cite{Zaa92}. They would compete with ZR singlets
\cite{Zha88}. Recent electronic structure results for (Nd,Sr)NiO$_2$
found using a combination of dynamical mean-field theory of correlated
electrons and band-structure methods indicate a remarkable
orbital-selective renormalization of the Ni 3d bands \cite{Leo20}.
However, for a NiO$_4$ plaque in NdNiO$_4$ crystal fields are quite
different and the orbital $xy$ is the first orbital below $x^2-y^2$
orbital occupied by one hole, while the $z^2$ orbital is at the bottom
of the orbital states {\cblue \cite{Wu20}.
This sequence of the orbital states at a Ni ion was recently confirmed
by quantum chemistry calculations} \cite{Kat20}. It is thus challenging
to ask to which $3d$ orbital doped holes will go in a doped material
and whether this could have any physical consequences.

The paper is organized as follows. In Sec. \ref{sec:ctm} we explain the
charge transfer model and its parameters. The methods to analyze finite
systems, i.e., exact diagonalization and the variational cluster
approximation (VCA) are introduced in Sec. \ref{sec:vca}. The numerical
results are presented and discussed in Sec. \ref{sec:res}. The paper is
summarized in Sec. \ref{sec:summa}. In the Appendix we analyze briefly
the consequences of the partial filling of $xy$ orbital on Ni ions and
conclude that the general conclusions concerning the possibility of the
participation of high-spin states in the ground state are not affected.

\section{The charge-transfer model}
\label{sec:ctm}

We introduce the multiband $d-p$ Hamiltonian for a NiO$_2$ plane where
we consider a 2D $2\times 2$ cluster shown in Fig. \ref{cluster} (with
periodic boundary conditions) which includes four orbitals per NiO$_2$
unit cell: two $e_g$ orbitals \mbox{$\{3z^2-r^2,x^2-y^2\}$} at each
Ni$^+$ ion and one bonding $2p_\sigma$ orbital (either $2p_x$ or $2p_y$)
at each oxygen ion in the 2D plane (12 is the total number of ions in
the periodic Ni$_4$O$_8$ cluster),
\begin{equation}
{\cal H}=
H_{dp}+H_{pp}+H_{\rm diag}+H_{\rm int}^d+H_{\rm int}^p.
\label{model}
\end{equation}
Here the first two terms in the Hamiltonian (\ref{model}) stand for the
kinetic energy: $H_{dp}$ includes the $d-p$ hybridization
$\propto t_{pd}$ and $H_{pp}$ includes the interoxygen $p-p$ hopping
$\propto t_{pp}$,
\begin{eqnarray}
H_{dp}&=&\sum_{\{m\alpha;j\nu\},\sigma}\left(t_{m\alpha;j\nu}
 \hat{d}^{\dagger}_{m\alpha,\sigma}\hat{p}_{j\nu,\sigma}^{} + {\rm H.c.}\right),\\
H_{pp}&=&\sum_{\{i\mu;j\nu\},\sigma}\left(t_{i\mu;j\nu}
 \hat{p}^{\dagger}_{i\mu,\sigma}\hat{p}_{j\nu,\sigma}^{} + {\rm H.c.}\right),
\end{eqnarray}
where $\hat{d}_{m\alpha,\sigma}^{\dagger}$
($\hat{p}_{j\nu,\sigma}^{\dagger}$) is the creation operator of an
electron at nickel site $m$ (oxygen site $i$) in an orbital $\alpha$.
At Ni ions $\alpha\in\{z,\bar{z}\}$, where $z$ and $\bar{z}$ stands for
$3z^2-r^2$ and $x^2-y^2$ orbitals, while at O ions $\nu\in\{x,y\}$
stands for $p_x$ and $p_y$ orbital, with up or down spin,
\mbox{$\sigma\in\{\uparrow,\downarrow\}$.} The model includes two $3d$
orbital states of $e_g$ symmetry at Ni$^+$ ions, and one $2p_{\sigma}$
bonding oxygen orbital state at oxygen ions in the $(a,b)$ plane, either
$2p_x$ for the Ni-O bond $\langle mi\rangle\parallel a$ or $2p_y$ for
the bond $\langle mi\rangle\parallel b$.
We study an isolated NiO$_2$ plane in NdNiO$_2$, so do not consider
$2p_z$ orbitals at apical oxygen positions.

In the following we will use shorthand notation, and instead of
$\{x^2-y^2,3z^2-r^2\}$ we shall write $\{(\bar{z}),(z)\}$ ---
this emphasizes the fact that $z$ axis is chosen as the quantization
axis for this $e_g$ orbital basis, while "()" brackets are used here
to distinguish two Ni($3d$) orbitals from O($2p_{\sigma}$) orbitals,
labeled as $\{x,y\}$. The elements of the matrices
$\{t_{m\alpha;j\nu}\}$ and $\{t_{i\mu;j\nu}\}$ are assumed to be
non-zero only for nearest neighbor nickel--oxygen $d-p$ pairs, and for
nearest neighbor oxygen--oxygen $p-p$ pairs. The next nearest neighbor
hopping is neglected. (The nonzero $\{t_{m\alpha;j\nu}\}$ and
$\{t_{i\mu;j\nu}\}$ elements are standard and are listed, e.g. in the
Appendix of Ref.~\cite{Ros15}).

The one-particle (level) energies are included in $H_{\rm diag}$, where
the onsite elements of kinetic energy are bare level energies and local
crystal fields. The cluster geometry and precise forms of different
terms are standard; these terms were introduced in the previous
realizations of the three-band $d-p$ model for CuO$_2$ planes
\cite{Zaa88} and for other transition metal oxides \cite{Ros15,Ros19}.
The diagonal part $H_{\rm diag}$ depends on electron number operators.
It takes into account the effects of local crystal fields and the
difference of reference orbital energies
(here we employ the hole notation, $\Delta>0$),
\begin{equation}
\Delta=\varepsilon_d-\varepsilon_p,
\label{Delta}
\end{equation}
between $d$ and $p$ orbitals (for bare orbital energies) where
$\varepsilon_d$ is the average energy of all $3d$ orbitals, i.e., the
reference energy before they split in the crystal field due to the
surrounding oxygens. We fix this reference energy for all $3d$ orbitals
to zero, $\varepsilon_d=0$, and use only $\Delta=-\varepsilon_p$ and
thus we write,
\begin{eqnarray}
H_{\rm diag}=\Delta\sum_{i;\mu=x,y,z;\sigma}
\hat{p}^\dagger_{i\mu,\sigma}\hat{p}_{i\mu,\sigma}^{}
+\Delta_{z^2}\sum_{m\sigma}\hat{n}_{m(z)\sigma}.
\label{Dz}
\end{eqnarray}
The first sum is restricted to oxygen sites $\{i\}$, while the second
one runs over nickel sites $\{m\}$. The crystal-field splitting strength
vector ($\Delta_{z^2}$) describes the splitting between two $e_g$
orbitals; here we assume $\Delta_{z^2}=0$ to find the lowest value of
$\Delta$ for which the ground state of the NiO$_2$ plane changes.
A finite value of $\Delta_{z^2}=2.0$ eV is assumed in the Appendix.
Note that the charge-transfer energy $\Delta$ in Eq. (\ref{Delta}) is
defined for bare levels.

\begin{table}[t!]
\caption{Parameters of the multiband model (\ref{model}) (all in eV)
used in ED calculations to model a NiO$_2$ plane. For the hopping
integrals we adopt the values from Refs. \cite{miz95,miz96} and include
oxygen-oxygen hopping elements in $H_{pp}$ given by $(pp\sigma)=0.6$ eV
(here we use the Slater notation \cite{Sla54}).
The $t_{2g}$ and $e_g$ orbital states are arbitrarily taken as
degenerate.
}
\begin{ruledtabular}
\begin{tabular}{ccccccccc}
paper & $(pd\sigma)$ & $t_{pp}$ & $\Delta$ &
$U_d$ & $J_{\rm H}$ & $U_p$ & $J_{\rm H}^p$ & $U_{dp}$  \\
  \hline
\cite{Jia20} & $1.5$ & 0.55 & 7.0 & 8.34 & 1.18 & 4.4 & 0.8 &   $0.0$   \\
this work    & $1.3$ & 0.55 & 7.0 & 8.34 & 1.18 & 4.4 & 0.8 & $0.0-1.0$ \\
\end{tabular}
\end{ruledtabular}
\label{tab:para}
\end{table}

In Eq. (\ref{model}) $H_{\rm int}^d$ and $H_{\rm int}^p$ stand for the
intraatomic Coulomb interactions at Ni and O ions, respectively. Local
interactions at nickel ions, ${\cal H}_{{\rm int}}^d$, are rotationally
invariant in the orbital space \citep{Ole83} and depend on three Racah
parameters \cite{Racah}. For systems where only one kind of orbitals
are involved, two Kanamori parameters suffice:
(i) intraorbital Coulomb interaction $U$ and
(ii) Hund's exchange $J_{H}$. Here for a pair of $e_{g}$ electrons
in different orbitals one finds $J_H=4B+C$
(a~similar expression can be also written for $t_{2g}$ electrons),
\begin{align}
{\cal H}_{{\rm int}}^d & =U_d\sum_{i\alpha}
\hat{n}_{i\alpha\uparrow}\hat{n}_{i\alpha\downarrow}
+J_{H}\sum_{i,\alpha\neq\beta}
\hat{d}_{i\alpha\uparrow}^{\dagger}\hat{d}_{i\alpha\downarrow}^{\dagger}
\hat{d}_{i\beta\downarrow}^{}\hat{d}_{i\beta\uparrow}^{}\nonumber \\
&+\sum_{i}\left[\left(U_d-\frac{5}{2}J_{H}\right)
\hat{n}_{i(z)}\hat{n}_{i(\bar{z})}
-2J_{H}\hat{\vec{S}}_{i(z)}^d\!\cdot\!\hat{\vec{S}}_{i(\bar{z})}^d\right]\!.
\end{align}
Interorbital Coulomb interactions $\propto\hat{n}_{i\alpha}\hat{n}_{i\beta}$
are expressed in terms of orbital electron density operators for a
pair %
\mbox{%
$\alpha<\beta$%
}, %
\mbox{%
$\hat{n}_{i\alpha}^{}\!=\sum_{\sigma}\hat{n}_{i\alpha\sigma}^{}\!=
\sum_{\sigma}\hat{d}_{i\alpha\sigma}^{\dagger}\hat{d}_{i\alpha\sigma}^ {}$%
}. Orbital spin operators %
\mbox{%
$\hat{\vec{S}}_{i\alpha}\equiv
\{\hat{S}_{i\alpha}^{x},\hat{S}_{i\alpha}^{y},\hat{S}_{i\alpha}^{z}\}$%
} appear in the Hund's exchange term, %
\mbox{%
$-2J_{H}\hat{\vec{S}}_{i\alpha}\!\cdot\!\hat{\vec{S}}_{i\beta}$.%
} In a Mott insulator, charge fluctuations are quenched and electrons
localize due to large energy of the lowest multiplet state,
$(U-3J_{H})\gg t$, associated with high-spin charge excitation and
determining the fundamental Mott gap. For the present Ni$^+$ ions,
Hund's exchange $J_H$ stabilizes high-spin excited states of two holes
per site, with spin $S=1$.

Local interactions at oxygen ions, ${\cal H}_{{\rm int}}^p$, are again
rotationally invariant in the orbital space \citep{Ole83} and depend on
two Kanamori parameters: (i) intraorbital Coulomb interaction $U_p$ and
(ii) Hund's exchange $J_p$ between a pair of $2p$ electrons in
different orbitals,
\begin{align}
{\cal H}_{{\rm int}}^p & =
U_p\sum_{i\mu}\hat{n}_{i\mu\uparrow}^p\hat{n}_{i\mu\downarrow}^p
+J_H^p\sum_{i,\mu\neq\nu}
\hat{p}_{i\mu\uparrow}^{\dagger}\hat{p}_{i\mu\downarrow}^{\dagger}
\hat{p}_{i\nu\downarrow}^{}\hat{p}_{i\nu\uparrow}^{}\nonumber \\
&+\sum_{i,\mu<\nu}\left[\left(U_p-\frac{5}{2}J_H^p\right)
\hat{n}_{i\mu}^p\hat{n}_{i\nu}^p
-2J_H^p\hat{\vec{S}}_{i\mu}^p\!\cdot\!\hat{\vec{S}}_{i\nu}^p\right]\!.
\end{align}
Interorbital Coulomb interactions $\propto\hat{n}_{i\mu}^p\hat{n}_{i\nu}^p$
are expressed in terms of orbital electron density operators for a
pair \mbox{$\mu<\nu$},
\mbox{
$\hat{n}_{i\mu}^{p}=\sum_{\sigma}\hat{n}_{i\mu\sigma}^{p}=
\sum_{\sigma}\hat{p}_{i\mu\sigma}^{\dagger}\hat{p}_{i\mu\sigma}^{}$}.
Orbital spin operators, %
\mbox{%
$\hat{\vec{S}}_{i\mu}^p\equiv
\{\hat{S}_{i\mu}^{px},\hat{S}_{i\mu}^{py},\hat{S}_{i\mu}^{pz}\}$%
}, appear in the Hund's exchange term, %
\mbox{%
$-2J_p\hat{\vec{S}}_{i\mu}\!\cdot\!\hat{\vec{S}}_{i\nu}$.%
} In a Mott insulator, charge fluctuations are quenched and electrons
localize due to large energy of the fundamental Mott gap,
$(U_p-3J_H^p)\gg t$, associated with high-spin charge excitation of
$p^2$ in case of $p^1$ (or $p^4$ in case of $p^5$) in the ground state
configuration. Hund's exchange $J_H^p$ stabilizes then high-spin $S=1$
states at oxygen sites with two holes.

The parameters of the charge-transfer model Eq. (\ref{model}) used in
the present paper are given in Table I. They are compared with the
parameters used in Ref. \cite{Jia20}. The only difference is somewhat
smaller $(pd\sigma)=1.3$ eV hybridization element which we consider to
be more realistic, see Table I.

\section{Numerical methods}
\label{sec:vca}

\subsection{Exact diagonalization}

We start with the impurity model and investigate the ground state of
(NiO$_4)^{6-}$ cluster \cite{Jia20} which corresponds to the
(NiO$_2^{3-}$) unit in the NiO$_2$ plane doped by one hole. This
cluster has a small basis of orbital states $\{(z),(\bar{z}),x,y\}$
and is diagonalized exactly for $N_{\uparrow}=N_{\downarrow}=1$ holes,
i.e., one $\uparrow$-spin and one $\downarrow$-spin hole.
Next we calculate the ground states of the model (\ref{model}) for the
(NiO$_2)_4$ cluster with periodic boundary conditions using the Lanczos
algorithm \cite{Koch} for the orbital basis $\{(z),(\bar{z}),x,y\}$ in
each unit cell. We start with the undoped system, i.e.. 4 holes in this
cluster, in the $N_{\uparrow}=N_{\downarrow}=2$-hole configuration.
Doped systems contain $N_{\uparrow}=3$, $N_{\downarrow}=2$ and
$N_{\uparrow}=N_{\downarrow}=3$ holes, respectively.

\subsection{Variational Cluster Approach}

We complement our exact diagonalization study by using the variational
cluster approach (VCA)~\cite{Pot03b}, which embeds the four-unit-cell
cluster into a larger system. This gives us some some effective access
to larger lattices, e.g. momentum-resolved one-particle spectral
densities or density of states. We use (Lanczos) exact diagonalization
to obtain the Green's function of a cluster $G_{\textrm{Cl}}$,
consisting of four Ni and six oxygen ions. The cluster self-energy
$\Sigma_{\textrm{Cl}}$ is then extracted from $G_{\textrm{Cl}}$ and
inserted into the one-particle Green's function $G$ of the thermodynamic
limit. The approximation consists of replacing the (unknown)
thermodynamic-limit self-energy by that of the cluster, i.e., by setting
$\Sigma\equiv\Sigma_{\textrm{Cl}}$.

According to the self-energy functional theory~\cite{Pot03a}, the
optimal cluster self-energy is the one optimizing the thermodynamic
grand potential,
\begin{align}\label{eq:omega_VCA}
\Omega = \Omega_{\textrm{Cl}}+\textrm{Tr} \ln G-\textrm{Tr}\ln G_{\textrm{Cl}}\;.
\end{align}
The parameters $\{\vec\tau^\prime\}$ that can be varied to find
optimized $\Sigma_{\vec\tau^\prime}$ are any one-particle parameters
used when solving the small cluster. We focus here on an overall
fictitious chemical potential $\mu'$ to ensure thermodynamic
consistency~\cite{Aic05} as well as a staggered antiferromagnetic field
$h'$ favoring N\'eel order. Neither has been seen to significantly
affect results reported here.

Static observables like spin or orbital densities as well as dynamic
quantities like the one-particle density of states of
$\mathbf{k}$-resolved one-particle spectra are then obtained from the
approximated Green's function $G$. The kinetic energy is thus fully
included while interaction effects are truncated to those accessible
to the directly solved cluster. A slight difficulty lies in the Coulomb
repulsion $U_{pd}$ between Ni and O ions, as this can connect orbitals
inside the cluster with others outside. In contrast to inter-cluster
hopping, such inter-cluster interactions cannot rigorously be included
in the VCA, but have to be truncated or approximated. We have here
allowed periodic boundary conditions for these interactions. While this
would likely be too crude an approximation if intersite interactions
are close to driving an ordered phase (e.g. a charge-density wave)
\cite{Aic04,Dag14,Adl19}, it has been shown good enough to stabilize
uniform charges in the study of a strongly related three-band model for
cuprates \cite{Arr09}. Since the oxygen ions have lower occupation in
our present work, the approximation can here be expected to be even
less critical.

\section{Ground state at increasing $\Delta$}
\label{sec:res}

\subsection{ Impurity model}
\label{sec:resa}

\begin{figure}[b!]%
\noindent\centering{}\includegraphics[width=1\columnwidth]{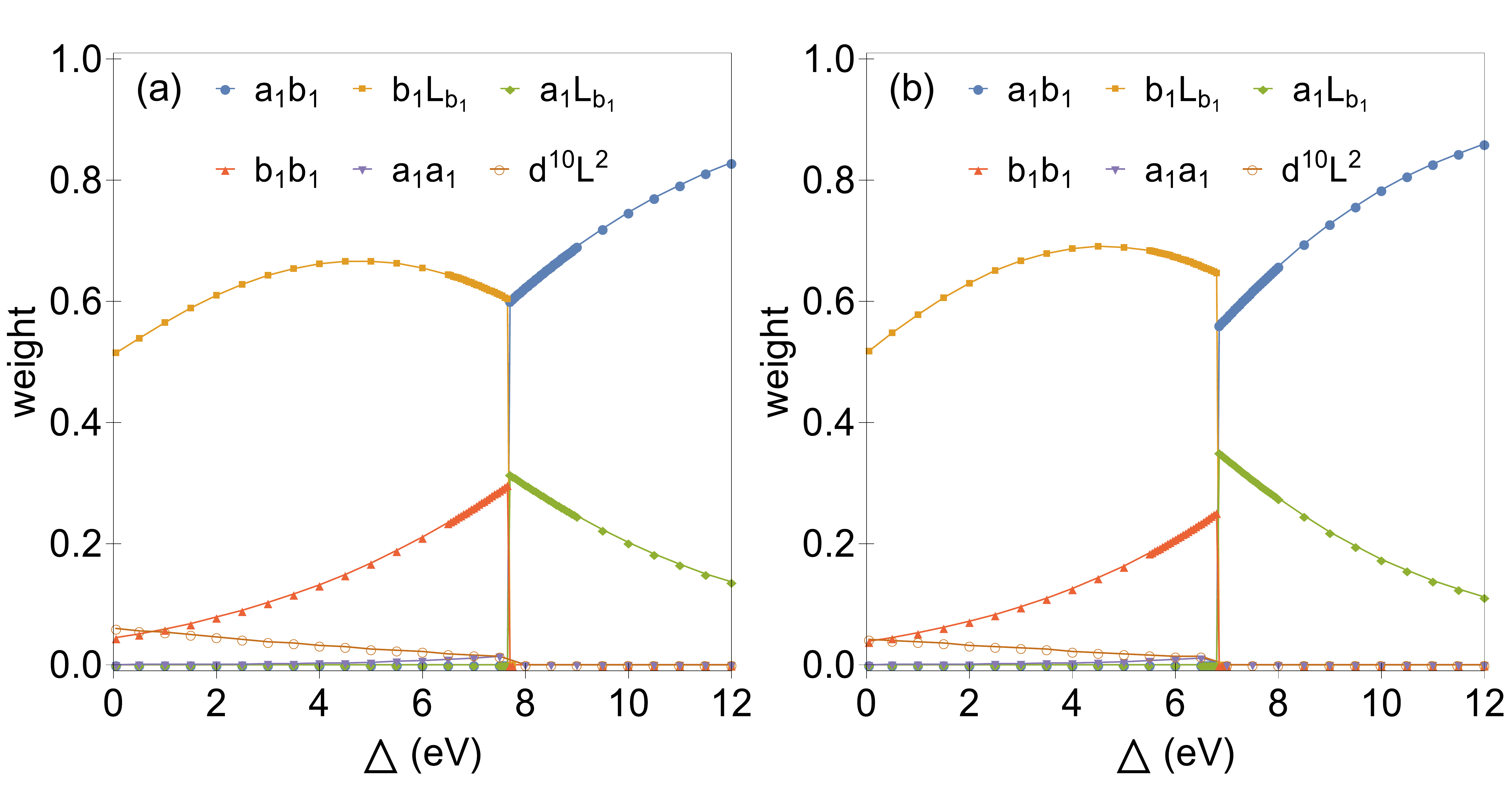}
\caption{
Weight distribution of different components in a single NiO$_2^{2-}$
cluster of a doped NiO$_2$ plane, with
$N_{\uparrow}=N_{\downarrow}=1$ hole, for two parameter sets of Table I:
(a) our calculation reproduces qualitatively the transition between two
ground state symmetries as in Fig. 4 in Ref. \cite{Jia20};
the transition is seen at somewhat lower $\Delta\simeq 7.7$ eV;
(b) somewhat enhanced triplet symmetry for finite $U_{dp}=1$ eV.}
\label{fig2}
\end{figure}

First we analyze the weight distribution in the ground state of the
doped (NiO$_2$)$^{2-}$ cluster with 2 holes per unit cell in
Ni($3d^{10}$) and O($2p^6$) states, i.e., for the filling by
\mbox{$N_{\uparrow}=N_{\downarrow}=1$} hole. In Fig. \ref{fig2}(a) we
see a rapid change from the singlet ($S=0$) to triplet ($S=1$) state
at $\Delta_c\simeq 7.7$ eV. This value is somewhat lower than
$\Delta_c=8.1$ eV given in Ref. \cite{Jia20} as we use here a somewhat
smaller hybridization, see Table I. But overall we find a very good
qualitative agreement with a distinct transition from the ZR singlet
state dominated by the $b_1L_{b_1}$ configuration to the $a_1b_1$ state,
where both holes are mainly at Ni ion in the high-spin configuration.
We observe that this transition is governed by the interplay between the
$d-p$ hybridization and Hund's exchange at Ni ion, $J_H$.

The ground state for $\Delta<\Delta_c$ consists of one hole occupying
the $|(\bar{z})\rangle$ and the oxygen state with the same symmetry
constructed out of four bonding $\{p_{i\sigma}^{\dagger}|0\rangle\}$
states, i.e.,
\begin{equation}
 p_{m\sigma}^{\dagger}=\frac{1}{2}\left(p_{1\sigma}^{\dagger}
-p_{2\sigma}^{\dagger}-p_{3\sigma}^{\dagger}+p_{4\sigma}^{\dagger}\right),
\end{equation}
and the ZR singlet at site $m$ is,
\begin{equation}
|\phi_m\rangle=\frac{1}{\sqrt{2}}\left(
  d_{m(\bar{z})\uparrow}^{\dagger}p_{m\downarrow}^{\dagger}
- d_{m(\bar{z})\downarrow}^{\dagger}p_{m\uparrow}^{\dagger}\right)
  \left|0\right\rangle.
\label{zr}
\end{equation}
The weight of the ZR singlet is shown in Fig. \ref{fig2} and labeled as
$b_1L_{b_1}$. At $\Delta>\Delta_c$ the above weight becomes negligible
and the largest weight is found instead for the
$|\psi_m\rangle=|a_1b_1\rangle$ state which is antisymmetric for spins
in $\{(z),(\bar{z})\}$ orbitals and labeled as $a_1b_1$.

Increasing intersite Coulomb repulsion to $U_{dp}=1$ eV moves the above
transition to a value of $\Delta_c$ roughly lowered by $U_{dp}$, in
agreement with the expectation from the mean field approximation.
Otherwise the weights of different configurations are almost the same,
so one finds that this intersite charge-charge repulsion influences the
energies of the two competing states and not their internal structure.

\subsection{Model of NiO$_2$ plane}

\begin{figure}[b!]
\noindent \centering{}\includegraphics[width=1\columnwidth]{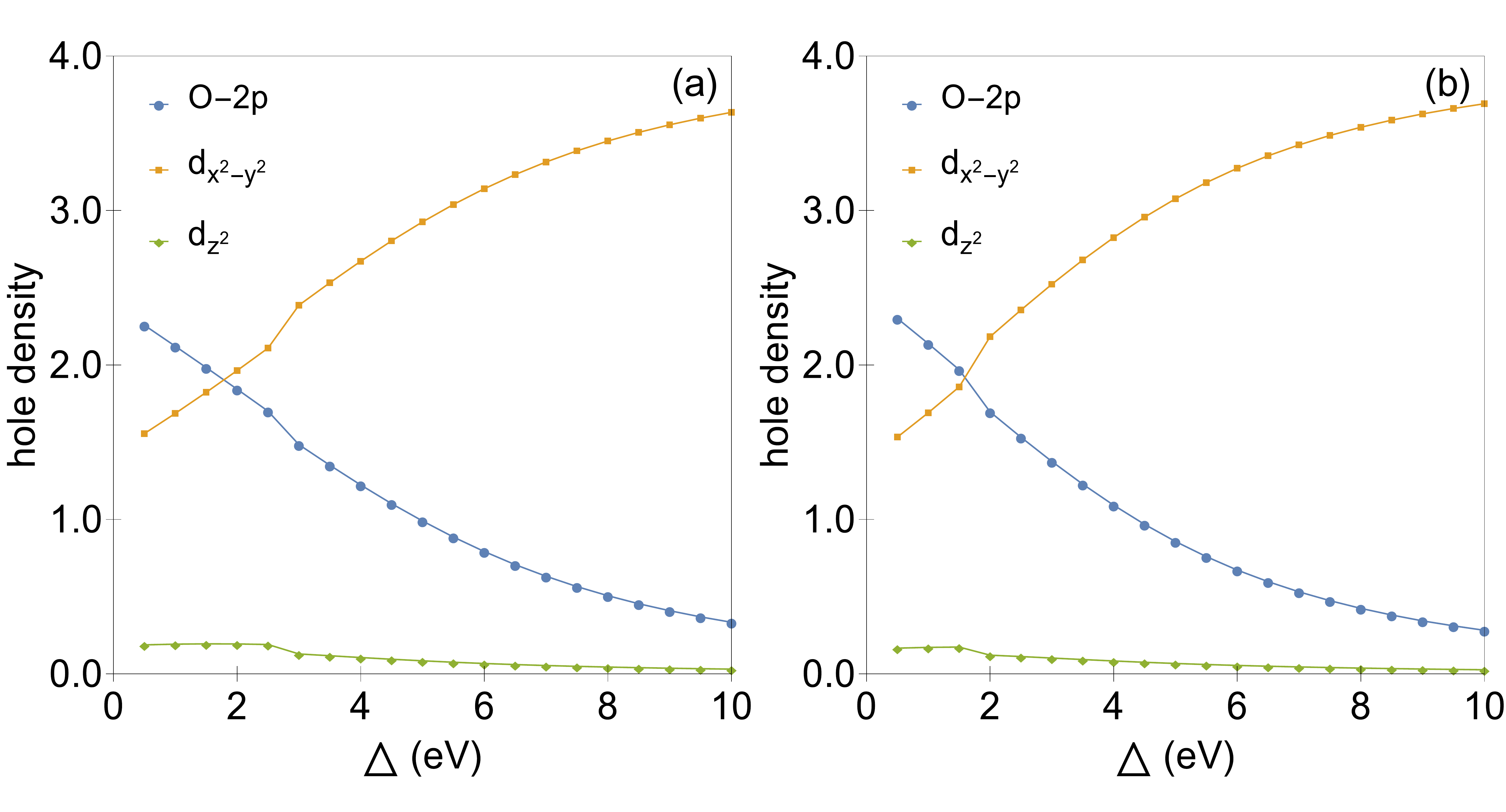}
\caption{
Hole densities as obtained in the undoped system Ni$_4$O$_8$ of Fig. 1,
$N_{\uparrow}=N_{\downarrow}=2$ holes, for increasing charge transfer
energy $\Delta$, and for:
(a) $U_{dp}=0$ and
(b) $U_{dp}=1.0$ eV.
Different lines show the hole densities in different orbitals:
oxygen $2p$ orbitals (blue), $d_{x^2- y^2}$ orbital (orange) and $d_{z^2}$
(green). }
\label{fig3}
\end{figure}

We next consider a cluster of $2\times 2$ unit cells (each containing
one Ni and two O ions) with periodic boundary conditions. The undoped
system corresponds then to one hole per Ni, i.e., to two holes with
spin up and down, $N_{\uparrow}=N_{\downarrow}=2$ holes within a
Ni$_4$O$_8$ cluster. Orbital-resolved densities are shown in
Fig.~\ref{fig3} and go from an almost even distribution between
$x^2-y^2$ and $2p$ orbitals at small $\Delta$ to nearly complete
localization on $x^2-y^2$ at large $\Delta$. Since there is on average
just one hole per one NiO$_2$ unit cell, there are only very few
configurations where two holes may interact by $U_{dp}$ and this
parameter does almost  not change the ground state at $U_{dp}=0$,
cf. Figs. \ref{fig3}(a) and \ref{fig3}(b).

Consider first the undoped system, {\cblue i.e., the one with one hole
per NiO$_2$ unit cell. The hole is then predominantly in the orbital of
$x^2-y^2$ symmetry.} The weights of these configurations
increase with increasing charge transfer energy $\Delta$ when the holes
hybridize more weakly with the surrounding oxygens. The many-hole wave
function may be written here as
\begin{eqnarray}
|\Phi_{22}\rangle&=&\left\{a_{(\bar{z})(\bar{z})(\bar{z})(\bar{z})}
\Pi_{m\in K_{\uparrow}}  d_{m(\bar{z})\uparrow}^{\dagger}
\Pi_{n\in L_{\downarrow}}d_{n(\bar{z})\downarrow}^{\dagger}
+... \right.
\nonumber\\
&+&\left.b_{({z})({z})({z})({z})}
\Pi_{m\in K_{\uparrow}}  d_{m(z)\uparrow}^{\dagger}
\Pi_{n\in L_{\downarrow}}d_{n(z)\downarrow}^{\dagger}
+...
\right\}|0\rangle, \nonumber\\
\label{real}
\end{eqnarray}
where $K_{\sigma}$ and $L_{\sigma}$ are the sets of Ni ions which are
occupied by $\sigma$-spin electrons, and
$a_{(\bar{z})(\bar{z})(\bar{z})(\bar{z})}$ and
$b_{({z})({z})({z})({z})}$ are the coefficients of two competing states.
For large charge-transfer energy $\Delta>6$ eV the holes concentrate
within the $|x^2-y^2\rangle$ state at each Ni site in the undoped
system. The hole density for the second $e_g$ orbital almost vanishes
and the high-spin states play no role.

\begin{figure}[t!]
\noindent \centering{}\includegraphics[width=.97\columnwidth]{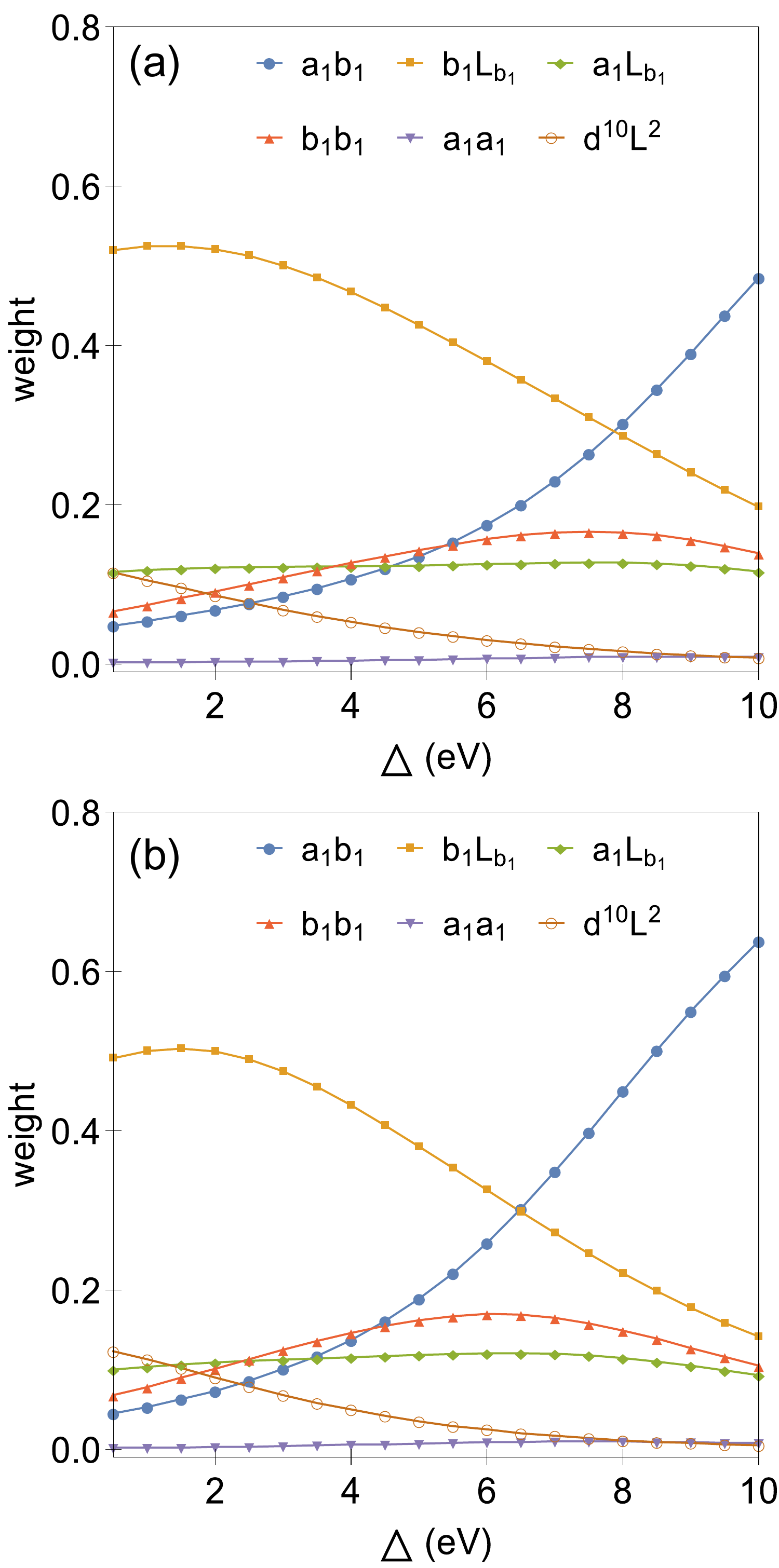}
\caption{The weights of various components (see legend) in the ground
state of the Ni$_4$O$_8$ cluster shown in Fig. \ref{cluster} as
obtained for increasing $\Delta$, $N_{\uparrow}=3$, and
$N_{\downarrow}=2$ for:
(a) $U_{dp}=0$ and
(b)~$U_{dp}=1.0$ eV.
The crossing point between the two ground state components with the
largest weights, $b_1L_{b_1}$ and $a_1b_1$, defines the crossover point
between the two types of the ground state. It is moved to a smaller
value of $\Delta$ when $U_{dp}=1.0$ eV.}
\label{fig4}
\end{figure}

Going to a doped system, there are many more configurations in the real
space and a very complex wave function similar to Eq. (\ref{real}) will
arise. Clearly, specifying the hole configurations in real
space is unpractical. Therefore we shall focus attention on the leading
two-hole configurations, such as these considered for an isolated
NiO$_4$ cluster, see Fig. \ref{fig2}. Thus for total 5 holes in the
Ni$_4$O$_8$ cluster, $N_{\uparrow}=3$ and $N_{\downarrow}=2$, we shall
have components in the total wave function which have two holes in the
states centered at one Ni site and otherwise one hole for the other
three Ni sites. For the particular Ni site $m$ with two holes we shall
write approximately the function centered at this site $m$ and two
holes with opposite spins in the both components,
\begin{eqnarray}
|\Phi_m\rangle\!&\approx&
a_m\frac{1}{\sqrt{2}}\left(
  d_{m(\bar{z})\uparrow}^{\dagger}p_{m0\downarrow}^{\dagger}\!
- d_{m(\bar{z})\downarrow}^{\dagger}p_{m0\uparrow}^{\dagger}\right)
\left|0\right\rangle\nonumber\\
&+&b_m\frac{1}{\sqrt{2}}\left(
 d_{m(z)\uparrow}^{\dagger}d_{m(\bar{z})\downarrow}^{\dagger}\!
+d_{m(z)\downarrow}^{\dagger}d_{m(\bar{z})\uparrow}^{\dagger}\!\right)
\left|0\right\rangle.
\label{Phim}
\end{eqnarray}
Here the first line is the ZR singlet and the second line stands for
the $|a_1b_1\rangle$ state considered before in Sec. \ref{sec:resa},
and $\{a_m,b_m\}$ are the probability amplitudes of each of the
elemental configurations. Of course, the ground state is more complex
and includes also other configurations. Taking $|a_m|^2$ and $|b_m|^2$
as the respective probabilities, we obtain the main local configurations
with their weights $\{b_1L_{b_1},a_1b_1\}$ displayed in Fig. \ref{fig4}.
The finite weights of other configurations displayed as well in Fig.
\ref{fig4} are obtained in a similar way numerically and demonstrate
that delocalization of holes over oxygens in the regime of small
$\Delta$. Analyzing these weight distributions shows that the character
of the wave function changes when the charge transfer energy increases
from small to large $\Delta$.

Figure~\ref{fig4} refers to the lowest doping achievable on a cluster
with four unit cells, namely one additional hole, i.e., $x=\frac14$
doped hole per Ni site, or $N_{\uparrow}=3$ and $N_{\downarrow}=2$. In
contrast to the single Ni ion results shown in Fig. \ref{fig2}, we no
longer see a sharp jump, but rather a gradual transition. This is due to
the transition from a rotationally invariant impurity to a translational
invariant lattice, where states of, e.g. $b_1$ and $a_1$ symmetry, are
allowed to hybridize. In agreement with this wave function, we observe
that the most important two configurations for the ground state are:
$|b_1L_{b_1}\rangle$ and $|a_1b_1\rangle$ states. They stand for the ZR
singlet state and for the high-spin $S=1$ state stabilized by Hund's
exchange when $\Delta>\Delta_c$. As observed for the impurity model in
Fig.~\ref{fig2} and as expected, charge-transfer energies
$\Delta\lesssim 7.0$ eV lead to a large weight in the ZR-singlet state,
while the triplet state localized at Ni dominates at large $\Delta$.
Again, non-local Coulomb repulsion $U_{dp}$ shifts this transition
precisely in the range of $\Delta\approx 6-8$ eV expected to apply to
nickelates, with $U_{dp}>0$ favoring the local triplet state already at
lower crystal fields.

\begin{figure}[t!]
\noindent \centering{}\includegraphics[width=.97\columnwidth]{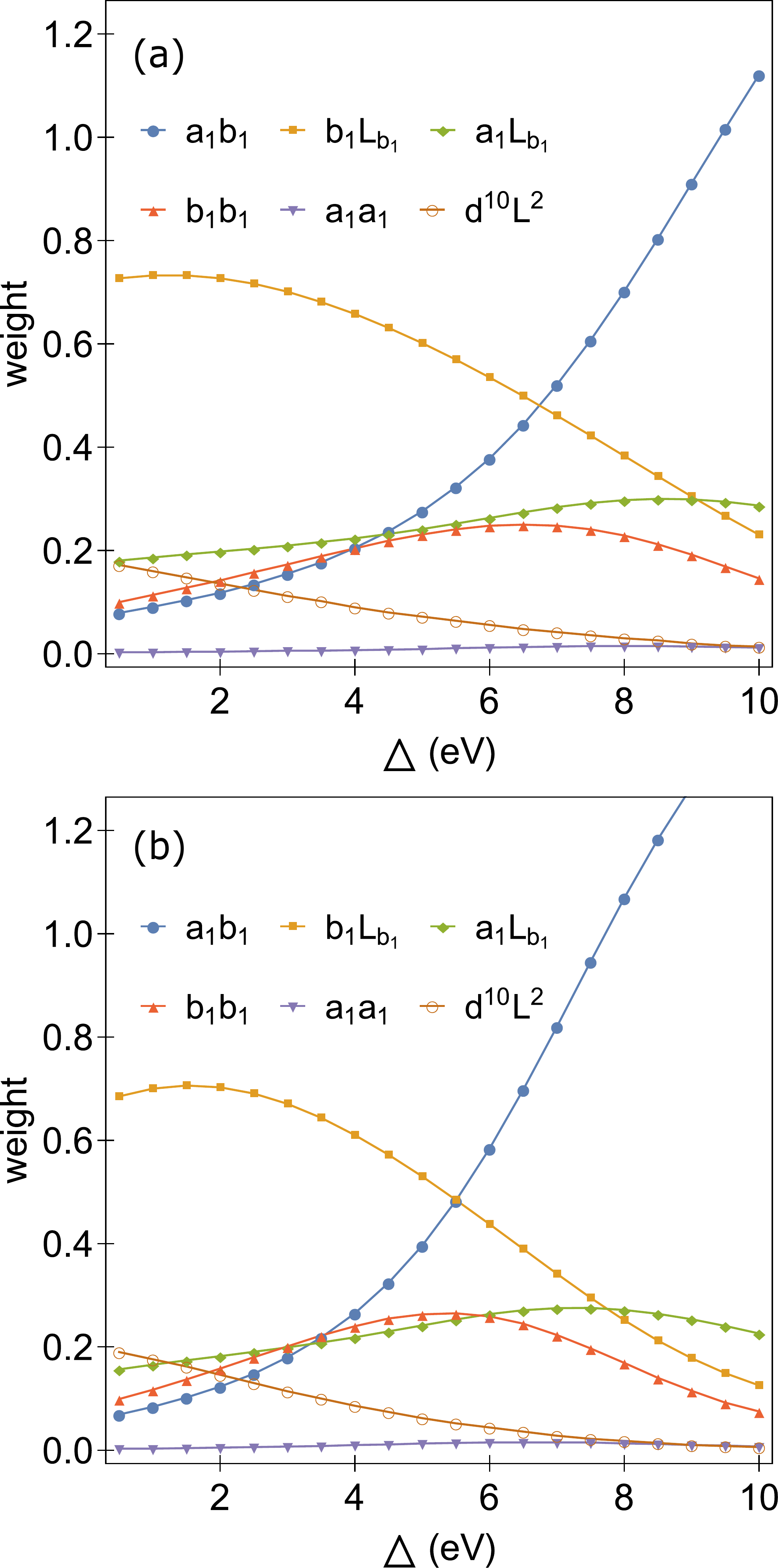}
\caption{The weights of the most important components in the ground
state of the Ni$_4$O$_8$ cluster shown in Fig. \ref{cluster} with one
or two holes at Ni sites, as obtained for increasing $\Delta$,
$N_{\uparrow}=N_{\downarrow}=3$, and for:
(a)~$U_{dp}=0$, and
(b) $U_{dp}=1.0$ eV.}
\label{fig5}
\end{figure}

As already apparent in the impurity model Fig.~\ref{fig2}, other
states in addition to the 'pure' ZR-singlet and onsite $S=1$ states,
more wave functions contribute, especially around the critical $\Delta
\approx 7.0$ eV. For instance, a 'ZR' variant of the high-spin state
contributes, where both holes have parallel spins (or are the third
component of the triplet, but one hole is located on $2p$ orbitals.
This state has rather constant weight over the whole range of $\Delta$,
so that some $3z^2-r^2$ character is expected for holes doped even into
a cuprate regime with $\Delta=3.0$ eV. This can be understood as due to
delocalization of the $2p$ hole over adjacent $3z^2-r^2$ orbitals.
Conversely, doubly occupied $x^2-y^2$ orbitals also have a relevant
weight for intermediate crystal fields $4.0\lesssim\Delta\lesssim 10.0$
eV. For a system still in the regime of predominant singlets
($\Delta\lesssim 7.0$ eV), the wave function is thus not necessarily
a ZR singlet familiar from cuprates, where one hole has clear $2p$
character, but can already have substantial 'Mott' character with
partly double occupied $x^2-y^2$ orbitals.

\begin{figure}[t!]
\noindent \centering{}\includegraphics[width=0.88\columnwidth]{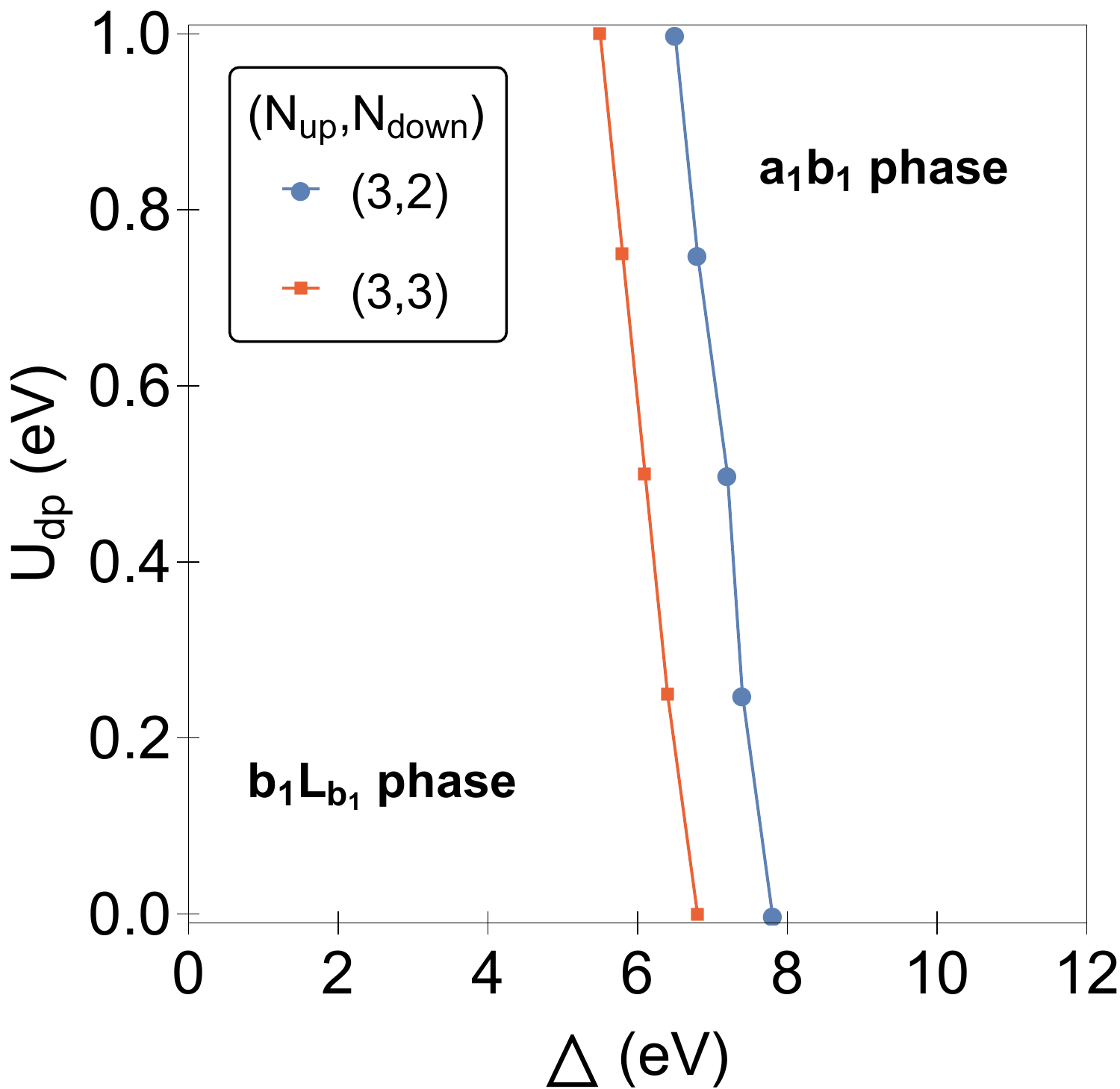}
\caption{Phase diagram arising from comparing the weight of the two
dominant ground state components for different values of $\Delta$ and
$U_{dp}$. Blue line for the Ni$_4$O$_8$ cluster doped by one hole
($N_{\uparrow}=3$, $N_{\downarrow}=2$) and red line for the Ni$_4$O$_8$
cluster doped by two holes ($N_{\uparrow}=N_{\downarrow}=3$).}
\label{fig6}
\end{figure}

When two holes are added to the Ni$_4$O$_8$ cluster, i.e.,
\mbox{$N_{\uparrow}=N_{\downarrow}=3$,} one finds locally once again
the low-spin ground state with ligand holes $|b_1L_{b_1}\rangle$ for
small values of $\Delta<5$ eV, see Fig. \ref{fig5}. This regime
corresponds to doped cuprates. For even smaller values of $\Delta$ one
finds stronger delocalization of holes onto the oxygen orbitals seen by
the component of the $|d^{10}L^2\rangle$ state. But for $\Delta\simeq 7$
eV as in nickelates \cite{Jia20}, this state plays no role and the
ground state is qualitatively different---it has the largest weight for
the high-spin state at Ni sites, $|a_1b_1\rangle$. This trend is
enhanced by the realistic finite value of $U_{dp}=1.0$ eV,
cf. Figs. \ref{fig5}(a) and \ref{fig5}(b).

The evolution of the ground state in the doped systems may be
qualitatively characterized by the transition from the ground state
dominated by the low-spin $|b_1L_{b_1}\rangle$ local states to the more
localized at Ni sites high-spin $|a_1b_1\rangle$ states, as shown in
Fig. \ref{fig6}. The delocalization over oxygen orbitals occurs easier
in the low doping regime $x=\frac14$, while for higher doping of
$x=\frac12$ the high-spin components dominate already for $\Delta<6.0$
eV, taking the realistic finite value of $U_{dp}=1.0$~eV.

\subsection{{\cblue Discussion:}
comparison between CuO$_2$ and NiO$_2$ plane}

\begin{figure}[t!]
\noindent
\centering{}\includegraphics[width=\columnwidth]{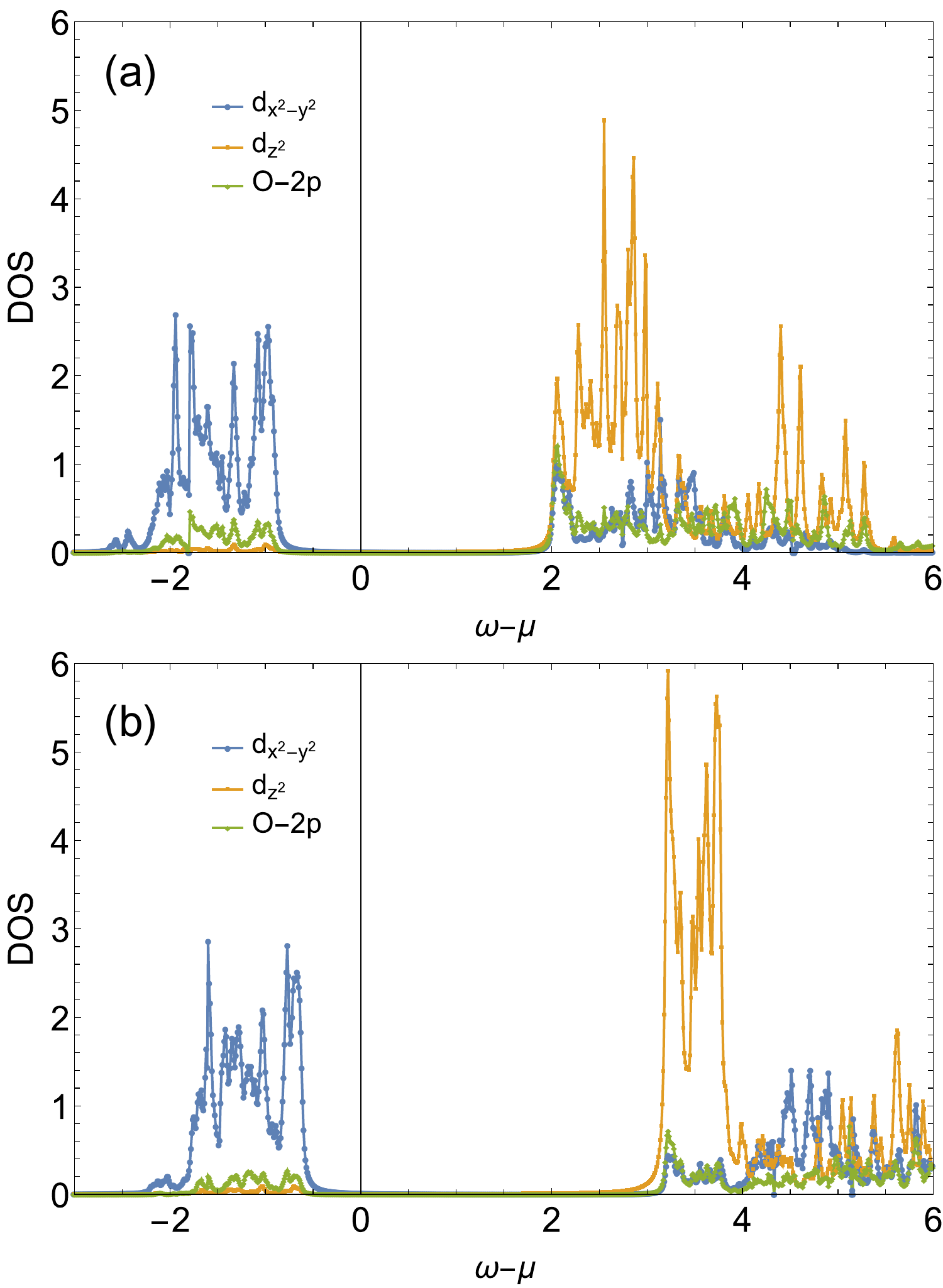}
\caption{Density of states for the reference undoped NiO$_2$ plane
obtained via VCA and resolved for each orbital:
blue $_{x^2-y^2}$, orange ${z^2}$ and (green) oxygen 2p orbital, for
$\Delta=7.0$ eV, and:
(a) $U_{dp}=0$, and
(b) $U_{dp}=1.0$ eV; the other parameters are given in Table I.
{\cblue The LHB at negative energies is predominantly occupied by
holes in $x^2-y^2$ orbitals. In the UHB} the occupation of ${z^2}$
states is enhanced while that of oxygen states is suppressed with
increasing $U_{dp}$.}
\label{fig7}
\end{figure}

{\cblue To illustrate the nature of the electronic states, we consider
the occupied and empty hole states now. As the (NiO$_2)^{3-}$ plane is
negatively charged, it is more convenient to use here the hole notation,
i.e., the occupied states at low energy in the lower Hubbard band (LHB)
are the states occupied by holes. In the undoped systems this means the
filling of one hole per each NiO$_2$ unit. First we consider such a
system and show that the undoped NiO$_2$ plane is insulating, see Fig.
\ref{fig7}. Whether or not this corresponds to the real situation in
NdNiO$_2$ is an open question---we suggest that this system is in a poor
metallic state after considerable self-doping. It increases the hole
concentration in NiO$_2$ planes beyond one hole per NiO$_2$ unit cell
when Nd ions have a smaller positive change than Nd$^{3+}$.}

We emphasize that due to the large value of $\Delta$ \cite{Jia20},
nickelates are in the Mott-Hubbard regime of the Zaanen-Sawatzky-Allen
diagram of correlated insulators \cite{ZSA}. This is also illustrated
by two Hubbard bands: the LHB and the upper Hubbard band (UHB),
separated by large Mott-Hubbard gap in the undoped system, see Fig.
\ref{fig7}. The LHB and the UHB have mainly the states at Ni sites,
with little oxygen admixture. The ground state contains predominantly
the holes within the $x^2-y^2$ orbitals in the LHB, while the $z^2$
states have large weight for the lower edge of the UHB. This weight is
enhanced for $U_{dp}=1.0$ eV, cf. Figs. \ref{fig7}(a) and \ref{fig7}(b).
Moreover, the character of the lowest UHB states changes with $U_{dp}$,
going from a somewhat ZR-singlet--like band (albeit with $z^2$
admixture and reduced $2p$ content) to an almost pure $z^2$ band.

\begin{figure}[t!]
\noindent \centering{}
\includegraphics[width=\columnwidth]{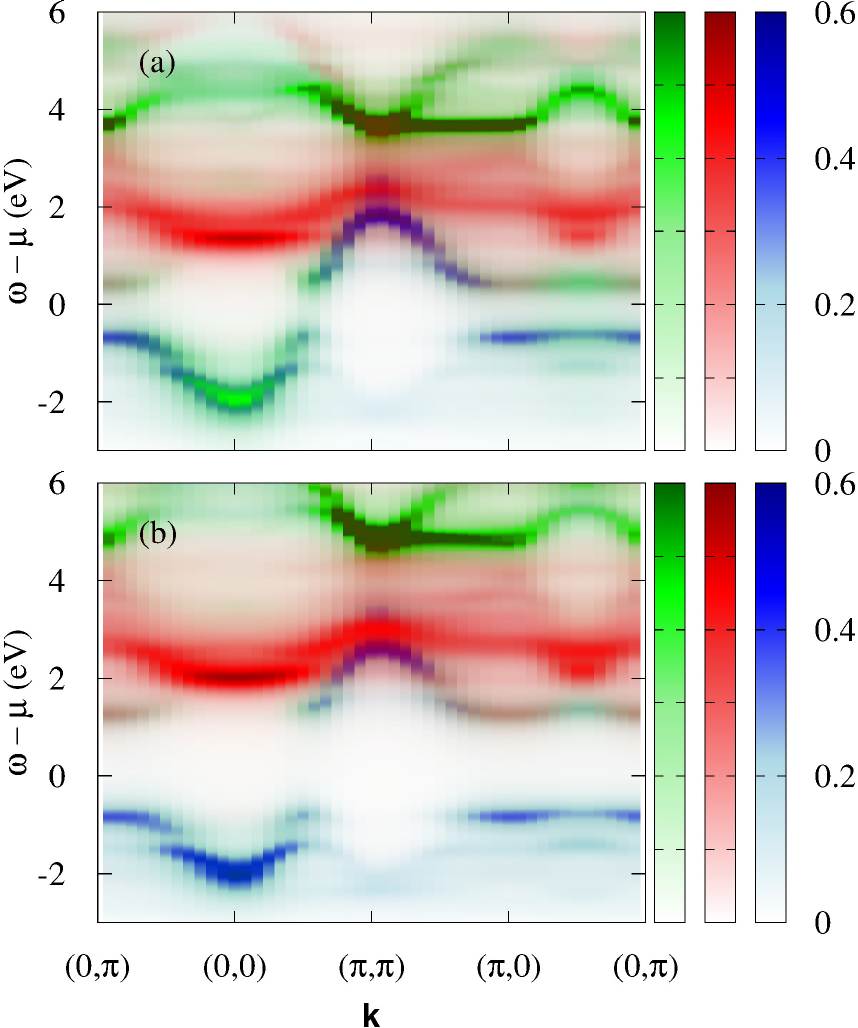}
\caption{The orbital-resolved and ${\bf k}$-resolved spectra for the
cuprate parameters with $\Delta=3.0$ eV along the high symmetry lines
in the 2D Brillouin zone for two values of $U_{dp}$:
(a) $U_{dp}=0$ and
\mbox{(b) $U_{dp}=1.0$ eV;} the other parameter as in Table I.
{\cblue Blue and red shading stands for $x^2-y^2$ and $z^2$ orbital
states at Ni ions, while green shading corresponds to oxygen states.}}
\label{fig8}
\end{figure}

To illustrate this point, let us first come back the the ZR-singlet
states arising in an analogous model applied to cuprates \cite{Arr09}.
With a lower value of $\Delta=3.0$ eV \cite{Arr09}, the CuO$_2$ plane
is in the charge-transfer regime. Hybridization between $x^2-y^2$
orbitals with oxygen $2p$ states is here naturally stronger, so that
oxygen content of occupied states rises. In the electronic structure
around the Fermi energy, shown in Fig.~\ref{fig8}, one find the lower
part of the UHB to be dominated by $x^2-y^2$ states. Nevertheless,
substantial oxygen-$p$ character is present, especially around
$(\pi/2,\pi/2)$ at the very lowest edges of the spectrum.

When $U_{dp}$ is increased from 0 to 1 eV, the gap increases and
orbital contributions to the bands are affected, but overall band
shapes and their order remain the same, see Fig.~\ref{fig8}(b). For
instance, one finds slight changes in the orbital character of the
lowest unoccupied states: in addition to $x^2-y^2$ and $2p$ orbitals,
even some $z^2$ weight is now present, especially around $(0,\pi)$ and
$(\pi,0)$. As can be seen in Fig.~\ref{fig4}, hole doping only induces
a very small triplet component for these parameters. The $z^2$ weight
seen in the spectra is thus more easily understood as arising from
delocalization of the ligand hole onto neighboring $z^2$ orbitals. When
$U_{dp}$ becomes relevant, it pushes more of the ligand holes into $z^2$
states. Despite this modification of orbital makeup, lowest states of
the UHB are naturally explained in terms of a single ZR-singlet band
for both $U_{dp}=0$ and 1 eV, while a $z^2$ band with a  minimum at
$(0,0)$ comes at higher energy.

\begin{figure}[t!]
\noindent \centering{}
\includegraphics[width=\columnwidth]{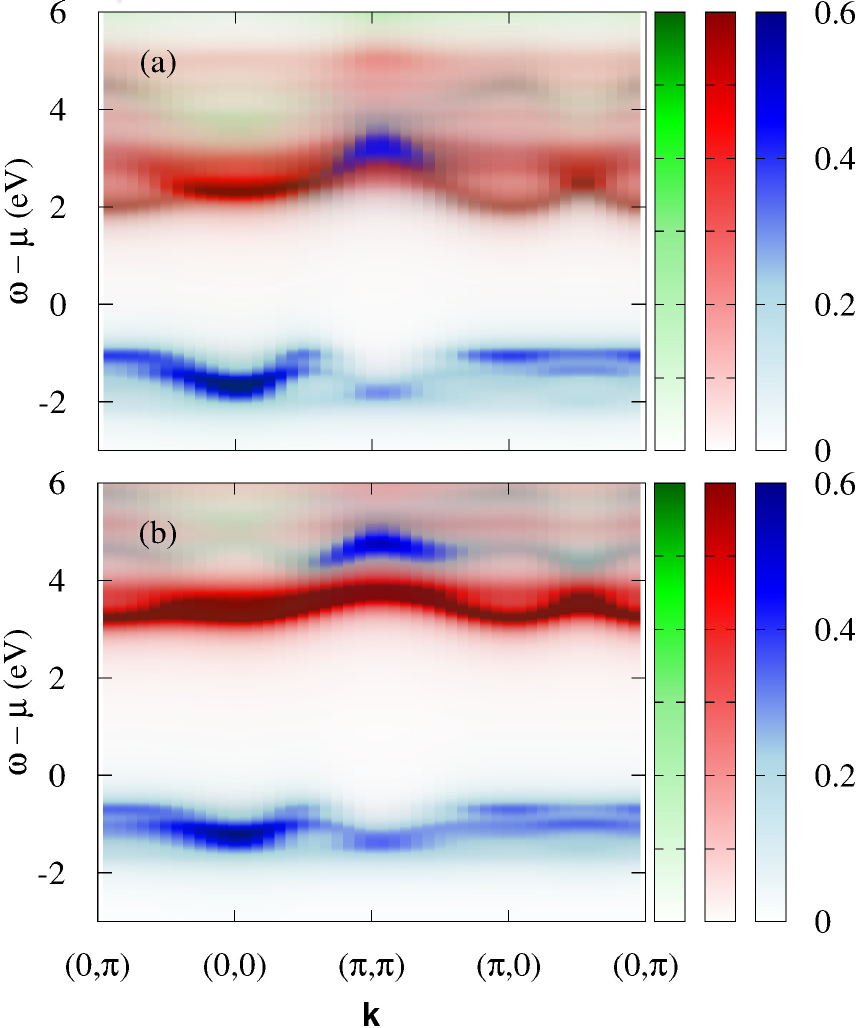}
\caption{The orbital and ${\bf k}$-resolved spectra for the nickelate
parameters (Table I) along the high symmetry lines in the 2D Brillouin
zone for: (a) $U_{dp}=0.0$ eV and (b) $U_{dp}=1.0$ eV;
the other parameter as in Table I.
{\cblue Blue and red shading stands for $x^2-y^2$ and $z^2$ states at
Ni ions, while green shading corresponds to oxygen states.}}
\label{fig9}
\end{figure}

Let us now come back to the nickelate regime with a larger $\Delta=7.0$
eV, for which $\mathbf{k}$-resolved spectra are shown in
Fig.~\ref{fig9}; they correspond to the density of states and the
Hubbard subbands of Fig.~\ref{fig7}. At first sight, one immediately
sees the stronger $z^2$ character near the lower edge of the UHB, both
without or with $U_{dp}$. The main oxygen bands, on the other hand, are
pushed to even higher energies outside the the energy window depicted.
The usual ZR-singlet picture does thus not apply, as one can also infer
from the sizable triplet component seen for $\Delta=7.0$ eV
in Fig.~\ref{fig4}.

For $U_{dp}=0$, however, the density of states in Fig.~\ref{fig7}(a)
clearly also shows some $2p$ and $x^2-y^2$ weight in the first
hole-doping states. The spectrum Fig.~\ref{fig9}(a) suggests the
corresponding lowest unoccupied band to be related to the ZR-singlet
band of the cuprate scenario Fig.~\ref{fig8}(b). The shape is similar,
with more pronounced minima at $(\pi,0)$ and $(0,\pi)$, but the
orbital makeup differs from cuprates, with much reduced $2p$ and much
increased $z^2$ character. A description of the lowest states in terms
of a ZR-singlet--like band appears still plausible, even though the
different orbital makeup and the resulting substantial triplet
admixture suggest that effective interaction parameters can be quite
different from those describing hole doping in cuprates.

At $U_{dp}=1\;\textrm{eV}$, finally, both the density of states in
Fig.~\ref{fig7}(b) and the $\mathbf{k}$-resolved spectrum in
Fig.~\ref{fig9}(b) show the lowest unoccupied states to be of almost
pure $z^2$ character. While some $2p$ and $x^2-y^2$ character is still
present, the system is now more easily understood as a 'classical'
Mott-Hubbard insulator, where the gap separates Hubbard subbands built
mainly by the orbital states at Ni ions. Accordingly, the onsite
triplet dominates over the ZR-singlet in Fig.~\ref{fig4}.

\section{Summary}
\label{sec:summa}

In summary, we have investigated the ground state of NiO$_2$ planes in
nickelate superconductors at increasing hole doping. First of all, we
find that the local high-spin states $|a_1b_1\rangle$ are important for
$\Delta\simeq 7.0$ eV, as expected for nickelate superconductors.
However, there is a crossover transition between the two regimes,
dominated by low-spin and high-spin states, when the charge transfer
energy increases.

Second, we would like to point out that $U_{dp}$ is an important
parameter to model the situation in doped NiO$_2$ planes. The Coulomb
repulsion between Ni and O sites influences the stability of low-spin
ZR states as this interaction is to a large extent missing in the
competing high-spin states. The value of $\Delta\simeq 7.0$ eV is in
the critical regime where the weight distribution and the nature of the
ground state depends in a subtle way on the value of $\Delta$.

Finally, as our third conclusion we wish to point out the increasing
admixture of high-spin states at Ni ions with increasing charge-transfer
energy $\Delta$ in the ground state of doped nickelates. We have shown
that the $z^2$ orbitals are the next to be occupied by doped holes in
case any extra splitting between the $z^2$ and the $x^2-y^2$ orbitals
is absent. In this way the holes concentrate at Ni sites and the
population of the $S=1$ states is increased. In nickelate films the
orbitals which are preferentially occupied are instead $xy$, but they
lead to the same global result that high-spin states arise locally. In
the Appendix we present arguments how the upper Hubbard band may change
when the $xy$ orbitals are filled instead by holes rather than the $z^2$
ones.

In the absence of electrostatic crystal field, the hybridization with
$2p$ orbitals indeed slightly favors the $z^2$ orbitals over the
$t_{2g}$ states, but the difference is not large. Electrostatic crystal
fields due to the planar geometry and missing apical oxygens can thus
easily tip the balance. However, the emerging scenario remains
equivalent: At $\Delta\approx 7.0$ eV and $U_{dp}=0$, hole doping still
leads to large singlet weight, while triplet states
(now involving the $xy$ orbitals) dominate for \mbox{$U_{dp}=1$ eV.}

In general, we find that hole doping distributes over the entire system,
in contrast to the onset of local $S=1$ states \cite{Zaa92}. This issue
is however still open and should be investigated further. At the moment
we can conclude that the contribution of high-spin states to the
electronic structure arises in any case, independently of which Ni
orbitals are occupied by a second hole in doped systems.

\acknowledgments

We thank Andres Greco, Krzysztof Ro\'sciszewski, and George A. Sawatzky
for very insightful discussions. We thank Ali Alavi for informing us
about the content of Ref. \cite{Kat20} prior to publication.
This research was supported in part by the National Science
Foundation under Grant No. NSF PHY-1748958.
T.~P. acknowledges \mbox{Development} and Promotion of Science
and Technology Talents Project (DPST). A.~M.~O. acknowledges
Narodowe Centrum Nauki (NCN, Poland) Project No. 2016/23/B/ST3/00839
and is grateful for the \mbox{Alexander von Humboldt} Foundation
\mbox{Fellowship}
\mbox{(Humboldt-Forschungspreis).}

\appendix*

\section{The model with $xy$ orbitals} \label{app}

There is some discussion concerning the order of the $d$ orbitals,
apart from the commonly accepted view that holes in the 'undoped'
reference system are preferentially found in $x^2-y^2$ orbitals.
Due to the planar cluster geometry (see Fig.~\ref{cluster}), one can
argue that the charge repulsion coming from the oxygen ions favors the
electron occupation of the $3z^2-r^2$ orbital and that the $xy$
orbital should come below the $x^2-y^2$ orbital, {\cblue as indeed
suggested by the crystal-field splittings \cite{Wu20}}. Such arguments
were also given for instance in Refs.~\cite{Dag20,Ros20} and are
supported by quantum chemistry calculations of hole levels~\cite{Kat20}.

On the other hand, one can argue that effective crystal-field splitting
is often determined by hybridization rather than electrostatic forces.
In this picture, putting a hole into the $3z^2-r^2$ orbital is
favorable because it can then delocalize better than the hole which
resides in the $xy$ orbital. Such effects resulting in the level order
opposite to the one expected from electrostatic charge have, e.g. been
observed in iridates with their extended $5d$ wave functions
\cite{Kim14}. In $a$ ($b$) direction, the corresponding hopping
parameters are $t_{pd}/2\simeq\sqrt{3}/2\simeq 0.866$ eV for the hopping
between the $3z^2-r^2$ orbital and the $p_x$ ($p_y$) orbital vs.
$(pd\pi)\simeq 0.75$ eV for hopping between $xy$ and $p_y$ ($p_y$).
Based on the analysis of Ref. \cite{Jia20a}, where hybridization effects
are found to dominate over point-charge effects, it has been argued
\cite{Jia20} that the $3z^2-r^2$ orbital should be considered
preferentially.

A~closer look at the results reported in Ref. \cite{Jia20a} indicates,
however, that the hybridization effects are not very large and in fact
the strongest in the 'charge-transfer' limit of a
smaller crystal field separating oxygen and transition metal ion. When
the hole becomes more localized onto the transition metal ion, impact
of hybridization with oxygen is reduced, so that the energy difference
between $xy$ and $3z^2-r^2$ shrinks. We thus investigate the aspect of
level ordering in this appendix.

To do so, we extend the model to include three $3d$ orbitals
\mbox{$\{x^2-y^2,3z^2-r^2,xy\}$} per each Ni ion as well as two ($p_x$
and $p_y$) at each oxygen, and study it using the VCA. We then have to
truncate the Hilbert space of the four-unit-cell cluster, but wish to
point out that a three--unit-cell cluster with the full Hilbert space
gave equivalent results. Since the $x^2-y^2$ orbital was nearly always
found to be half filled, we restrict possible states to these with at
least two holes in these orbitals in the cluster shown in Fig.
\ref{cluster}. In other words, we allow at most two of the four
$x^2-y^2$ orbitals to be completely filled.

Parameters referring to the orbitals included in the model with a
smaller basis set of the main text, as summarized in
Tab.~\ref{tab:para}, are kept here. In addition, Hund's exchange
coupling $J_H^p=0.8$~eV on oxygen ions as well as corresponding
interorbital $U_p^{'}=U_p-2J_H^p$ was introduced. Additional hoppings
were chosen following Ref.~\cite{Jia20a}: $t_{pp'}=-0.35$~eV
between nearest neighbor parallel oxygen orbitals and
$t_{xy}=t_{pd}/2=0.65$~eV connecting the $xy$ orbital to oxygen.
For the signs resulting from orbital phases, see~Ref. \cite{Jia20a}.

\begin{figure}[t!]
\includegraphics[width=\columnwidth]{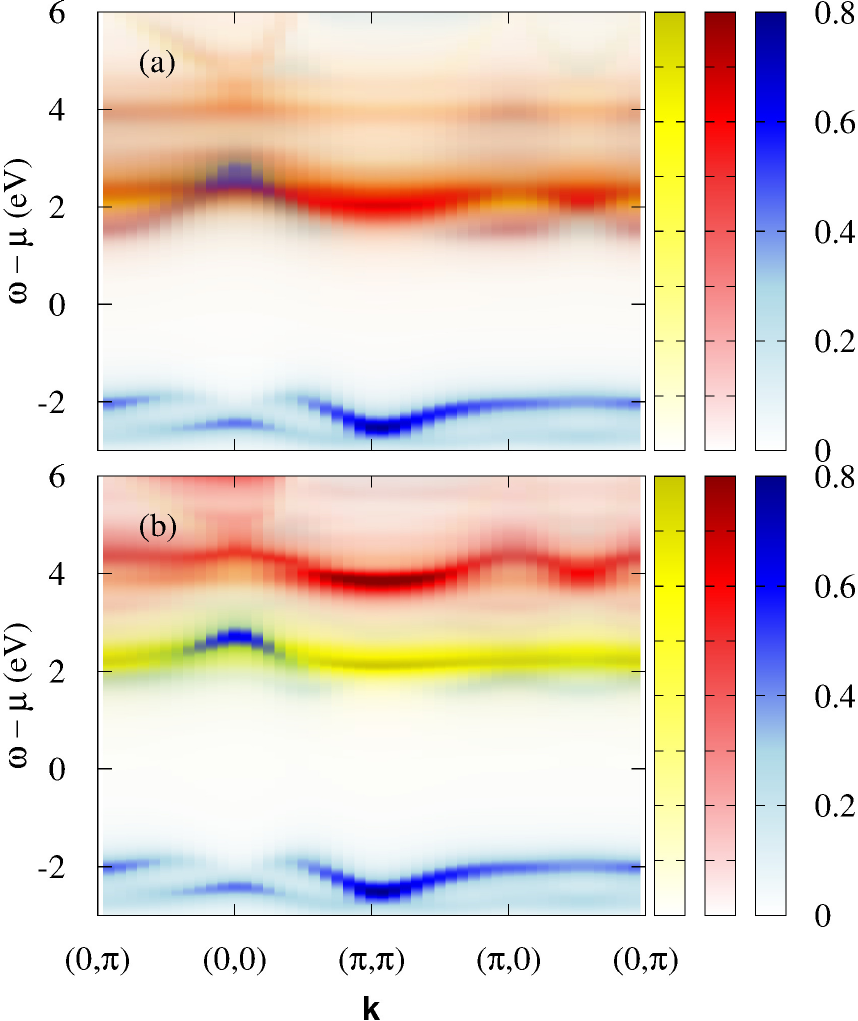}
\caption{$\mathbf{k}$-resolved one-particle spectra obtained for the
7-orbital model which includes three \mbox{$\{x^2-y^2,3z^2-r^2,xy\}$}
orbitals per Ni site and two $\{p_x,p_y\}$ orbitals at oxygen sites in
NiO$_2$ unit cell of Fig. \ref{cluster}. In (a) there is no explicit
crystal field between $x^2-y^2$ and \mbox{$3z^2-r^2$} orbitals
(\ref{Dz}), so that the $xy$ orbital lies slightly above the lowest
$3z^2-r^2$ states. In (b) the crystal field $\Delta_{z^2}=2.0$~eV, see
Eq. (\ref{Dz}), rises the $3z^2-r^2$ orbital, but the basic scenario of
holes entering a ZR-singlet band remains intact. The parameters as in
Tab.~\ref{tab:para}, and $\Delta=7.0$ eV, $U_{pd}=0$. Yellow shading
refers to the $xy$, red to $3z^2-r^2$, and blue to $x^2-y^2$ orbitals.
\label{fig10}}
\end{figure}

Without any static point-charge fields, the $xy$ orbital is indeed found
to be above the $3z^2-r^2$ orbital (in the hole notation), so that one
would encounter the latter first when hole doping the compound. {\cblue
Below we shall follow here the first principle calculations which give
the $xy$ orbitals as the energetically closest ones to $x^2-y^2$
\cite{Wu20}. If this is not assumed,} for $U_{dp}=0$,
holes enter first a band composed of a mixture of $x^2-y^2$,
$3z^2-r^2$, and $2p$ orbitals, see Fig.~\ref{fig10}(a). $3z^2-r^2$
states are mixed into this band, as seen before in Fig.~\ref{fig9}(a),
while $xy$ states come at slightly higher energies and do not mix
noticeably with the lowest hole band.

However, the splitting between the $3d$ states, \mbox{$3z^2-r^2$} and
$xy$, is very small indeed and moderate explicit crystal fields
\begin{align}
  H_{\Delta_{z^2}} = \Delta_{z^2} \sum_m n_{m,z^2}\;,
\end{align}
with the sum going over all Ni ions and $n_{m,z^2}$ referring to the
hole density in the $3z^2-r^2$ orbital, could easily reverse their
order. Such a situation is shown in Fig.~\ref{fig10}(b) for a moderately
large \mbox{$\Delta_{z^2}=2.0$~eV,} which is in line with recent
quantum-chemistry results~\cite{Kat20}. Nevertheless, conclusions from
the main text remain valid: for $U_{dp}=0$, lowest hole-doping states
are composed on $x^2-y^2$ and $2p$ states. In fact, their lower
$3z^2-r^2$-population and nearly absent $xy$ character makes this band
rather more similar to the ZR-singlet band known from cuprates.

\begin{figure}[t!]
\includegraphics[width=\columnwidth]{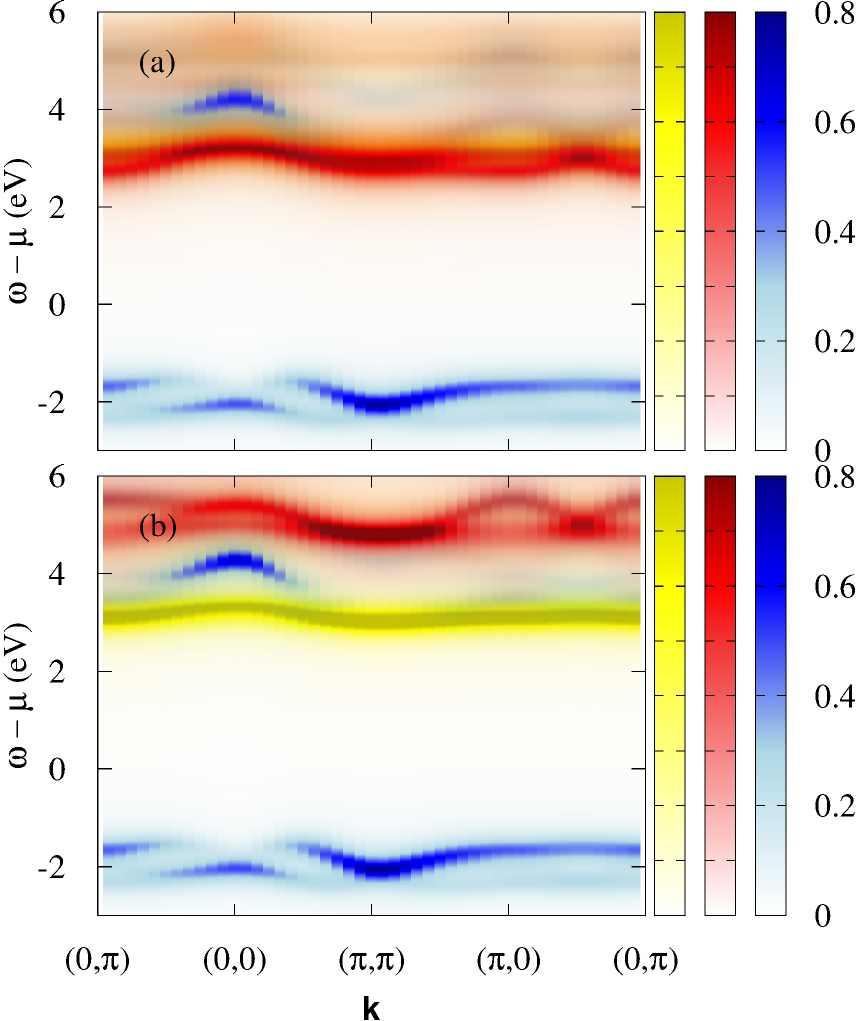}
\caption{As in Fig.~\ref{fig10}, but for $U_{pd}=1.0$~eV.
In (a) without crystal field, the lowest hole-doped states are now
$3z^2-r^2$, while they have $xy$ character when $\Delta_{z^2}=2.0$~eV
(\ref{Dz}) in (b).
{\cblue The color convention is the same as in Fig. \ref{fig10}.}
\label{fig11}}
\end{figure}

Once $U_{dp} =1\;\textrm{eV}$, Mott-Hubbard gap increases (Fig.
\ref{fig11}) and doped holes hardly go into the $x^2-y^2$ orbital,
regardless of whether $\Delta_{z^2}=0$ or $\Delta_{z^2}=2$ eV. For
$\Delta_{z^2}=0$, the lowest hole-doping states are now of almost
exclusively $3z^2-r^2$ character, see Fig.~\ref{fig11}(a). Once
\mbox{$\Delta_{z^2}=2.0$~eV}, the crystal field pushes the $3z^2-r^2$
states towards higher energies and holes enter a nearly pure $xy$ band
instead, see Fig.~\ref{fig11}(b).

We thus conclude that the $xy$ orbital is likely to be relevant to
doped NiO$_2$ planes, but that this does not affect our results:
for $U_{pd}=0$, a band with robust $x^2-y^2$ character orbital hosts
the lowest hole-doping states regardless of which orbital comes next,
see Fig.~\ref{fig10}. For $U_{pd}=1.0$~eV, conversely, holes enter
preferentially either $xy$ or $3z^2-r^2$ orbital, depending on their
relative crystal field, see Fig.~\ref{fig11}, and high-spin $S=1$
states form locally.

Since the $xy$ orbital bonds to the orbitals orthogonal to those
coupling with both the $x^2-y^2$ and $z^2$ orbitals, $x^2-y^2$ states
see less oxygen-mediated hybridization with $xy$ that with $z^2$ states.
The ZR-singlet--like band thus remains more clearly separated from the
$xy$ band than from the $3z^2-r^2$ states. In the full system,
however, the 'other' orbitals in addition to $x^2-y^2$ have been shown
to hybridize strongly with rare-earth states and also with each
other~\cite{Gu20}. This clear separation of $xy$ and $x^2-y^2$ orbitals
may thus be an artifact of our model, while the stronger mixing of
$x^2-y^2$-states with a second band ---as discussed in the main text---
may be more realistic~\cite{Hep20}.


%

\end{document}